\documentstyle [12pt] {article}
\title{DETERMINATION OF THE MINIMAL LENGTH BY MICROSCOPIC BLACK HOLE TEMPERATURE}

\author{Vladan Pankovi\'c$^{\ast,\sharp}$\\
$^\ast$Department of Physics, Faculty of Sciences, 21000 Novi
Sad,\\ Trg Dositeja Obradovi\'ca 4. , Serbia, vdpan@neobee.net \\
$^\sharp$Gimnazija, 22320 Indjija, Trg Slobode 2a, Serbia \\}

\date {}
\begin{document}
\maketitle

\vspace {0.5cm} PACS number: 04.70.Dy \vspace {0.5cm}

\begin {abstract}
Generalizing results of our previous work (where classical kinetic
energy has been used) in this work (where ultra-relativistic
kinetic energy is used) we suggest an original variant of the
determination of minimal length (corresponding to Plank length) by
formation of a microscopic (tiny) black hole. Like to some
previous authors we use Heisenberg coordinate-momentum uncertainty
relation, on the one hand. But, instead of metric fluctuation
(obtained by second derivative in Einstein equations) that
generalizes uncertainty relation by an additional term, used by
previous authors, we use Hawking temperature of the black hole and
standard Heisenberg coordinate-momentum uncertainty relation.
\end {abstract}

\vspace{1.2cm}

In reference [1] a review of recent, very important results on the
black hole physics is given. Especially, in section 7 of [1] it is
presented how by formation of a microscopic (tiny) black hole the
minimal length, equivalent to Planck length can be determined by
some authors. Simply speaking, by measurement of the coordinate of
a quantum system, or, simply speaking, a particle, standard,
Heisenberg coordinate-momentum uncertainty relation must be
satisfied. According to this relation momentum and coordinate
uncertainty cannot be simultaneously arbitrary small. But, by
arbitrary accurate measurement of coordinate corresponding
coordinate uncertainty can be arbitrary small. For this reason
standard Heisenberg coordinate-momentum uncertainty relation does
not predict non-zero minimal length. However, coordinate
measurement, roughly speaking, considers an interaction of given
particle with a measuring particle at a corresponding distance
(length). If given distance, by sufficiently accurate coordinate
measurement, becomes equal or smaller than Schwarzschild radius
for measured and measuring particle, then a microscopic (tiny)
Schwarzschild black hole will appear. It, according to Einstein
equations, will cause additional, nonstandard, metric fluctuation
or uncertainty, whose value is proportional to product of the
particle momentum and quadrate of Planck length. Or, there is a
generalization of the coordinate-momentum uncertainty relation so
that minimal length is equivalent to Plank length (smaller length
cannot be stablely measurable since corresponding measuring
procedure causes microscopic (tiny) black hole formation and
non-removable metric fluctuations).

In this work, generalizing results of our previous work [2] (where
classical kinetic energy has been used) we shall suggest (by use
of more accurate, ultra-relativistic kinetic energy) an original
variant of the determination of minimal length (corresponding to
Plank length) by formation of a microscopic (tiny) black hole.
Like to some previous authors we shall use Heisenberg
coordinate-momentum uncertainty relation, on the one hand. But,
instead of metric fluctuation (obtained by second derivative in
Einstein equations) that generalizes uncertainty relation by an
additional term, used by previous authors, we shall use Hawking
temperature of the black hole and standard Heisenberg
coordinate-momentum uncertainty relation.

We shall start from well-known, standard, Heisenberg
momentum-coordinate uncertainty relation
\begin {equation}
       \Delta p \Delta x \geq \frac {\hbar}{2}
\end {equation}
where, for measured quantum system, i.e. particle, $\Delta p$
represents the momentum uncertainty (or, standard deviation),
$\Delta x$ - coordinate uncertainty (or, standard deviation), and
$\hbar $ - reduced Planck constant. As it has been mentioned
previously, given uncertainty relation states that momentum
uncertainty and coordinate uncertainty cannot be simultaneously
arbitrary small. But given relation does not forbid such arbitrary
accurate measurement of coordinate within which coordinate
uncertainty can be arbitrary small. It implies possibility of
infinity small length. Namely, by physically relevant measurement
of an observable, e.g. coordinate, its average value cannot be
smaller than its standard deviation. But, since in sufficiently
accurate coordinate measurement coordinate uncertainty can be
arbitrary small, arbitrary small coordinate, i.e. length can be
physically relevant.

Uncertainty relation (1) implies
\begin {equation}
        \Delta p^{2}\simeq <p^{2}> - <p>^{2}\hspace {0.3cm}\geq \frac {\hbar^{2}}{4 \Delta x^{2}}
\end {equation}
and, further,
\begin {equation}
        <p^{2}> \hspace {0.3cm} \geq \frac {\hbar^{2}}{4 \Delta x^{2}} .
\end {equation}
where $<>$ denotes average value of corresponding observable.

For an ultra-relativistic quantum system, for which average value
of the total energy $<E>$ is practically equivalent to average
value of the kinetic energy $<E_{k}>$ or to $<p^{2}>^{\frac
{1}{2}}c$ where c represents the speed of light, (3) simply
implies
\begin {equation}
        \Delta x  \geq  \frac {\hbar  c}{2<E_{k}>}
\end {equation}

Suppose that for sufficiently accurate measurement of the
coordinate of particle, i.e. for sufficiently small $\Delta x$,
given particle becomes completely a Schwarzschild black hole. Then
average kinetic energy of the particle must be equivalent to
thermal energy of the black hole. (Precisely speaking, since
coordinate and momentum represent non-commutative observables
then, according to standard quantum mechanical propositions,
accurate measurement of coordinate causes significant statistical
fluctuations of the momentum. Also, since, in ultra-relativistic
case, momentum is practically proportional to kinetic and total
energy, given accurate coordinate measurement causes significant
statistical fluctuation of the kinetic energy too. For this reason
it seems quite naturally that given statistically averaged kinetic
energy of the particle be considered as the thermal energy, i.e.
energy of the statistical fluctuations of the black hole.) This
thermal energy corresponding to chaotic oscillations at black hole
horizon surface can be presented as $2(\frac {kT}{2}) = kT$ where
2 refers on two degree of freedom at horizon surface, while $k$
represents Boltzmann constant and $T$ - Hawking temperature. It
yields
\begin {equation}
        \Delta x  \geq  \frac {\hbar  c}{2kT}     .
\end {equation}
As it is well-known, Hawking temperature for Schwarzschild black
hole with mass $m$ equals
\begin {equation}
         T = \frac {\hbar c^{3}}{8k\pi Gm}
\end {equation}
where $G$ represents the gravitational constant. Then,
introduction of (6) in (5) yields
\begin {equation}
       \Delta x \geq  \frac{4\pi Gm}{c^{2}}= \frac{4\pi Gmc}{c^{3}}= 4 \pi (\frac{G\hbar}{c^{3}}) \frac {2\pi}{\lambda}=
       8\pi^{2} (\frac {G\hbar}{c^{3}}) \frac {1}{\lambda}= 8\pi^{2} L^{2}_{P}\frac {1}{\lambda}
\end {equation}
where $\lambda=\frac {2\pi \hbar}{mc}$ represents corresponding de
Broglie wave length while$ L_{P}  = (\frac {G\hbar}{c})^{\frac
{1}{2}}$ represents Planck length. Since, in given measurement
procedure, $\Delta x$ is practically equivalent to de Broglie wave
length, (7) implies
\begin {equation}
    \Delta x^{2}\geq 8\pi^{2}L^{2}_{P}
\end {equation}
and finally
\begin {equation}
   \Delta x \geq 8^{\frac {1}{2}}\pi L_{P}\sim L_{P}      .
\end {equation}

In this way we did an original variant of the determination of
minimal length (corresponding to Plank length) by formation of a
microscopic (tiny) black hole. Like to some previous authors we
use Heisenberg coordinate-momentum uncertainty relation, on the
one hand. But, instead of metric fluctuation (obtained by second
derivative in Einstein equations) that generalizes uncertainty
relation by an additional term, used by previous authors, we used
Hawking temperature of black hole and standard Heisenberg
coordinate-momentum uncertainty relation.

\vspace{2.5cm}

{\Large \bf References}

\begin {itemize}

\item [[1]] S. Hossenfelder, {\it What black holes can teach us}, hep-ph/0412265 and references therein
\item [[2]] V. Pankovic, {\it How we can do a microscopic (tiny) black hole and determine minimal length}, gr-qc/0901.3004

\end {itemize}

\end {document}